\documentclass{article}
\usepackage[utf8]{inputenc}
\usepackage{authblk}
\usepackage{array}
\usepackage[hang,flushmargin]{footmisc}
\usepackage[a4paper, margin=1in]{geometry}
\usepackage{indentfirst}
\usepackage[T1]{fontenc}
\usepackage{titlesec}
\usepackage[round, authoryear]{natbib}
\author{\vspace{-5ex}}
\date{\vspace{-5ex}}
\usepackage{setspace}
\usepackage{bm, amsmath, bbm}
\usepackage{graphicx}
\usepackage[hidelinks]{hyperref}
\usepackage{blindtext}
\usepackage{xcolor} 
\usepackage{soul}
\newcommand\mathbb[1]{\mathbbm{#1}}

\usepackage{fancyhdr}
\usepackage{lineno}
\usepackage{booktabs}
\usepackage{longtable}
\usepackage{multirow}
\usepackage{wrapfig}
\usepackage{float}
\usepackage{colortbl}
\usepackage{threeparttablex}

\begin{document}

\title{Network knockoffs: controlling false discovery in dyadic space}
\maketitle

\begin{center}
\Large{Justin J. \textsc{Van Ee}$^{1,2}$} 
\let\thefootnote\relax\footnote{\baselineskip=10pt 1, 
Department of Statistics and Data Sciences, The University of Texas at Austin, Austin, TX}\textrm{\Large{, }}
\Large{Yoichiro \textsc{Kanno}$^{2}$} 
\let\thefootnote\relax\footnote{\baselineskip=10pt 2, 
Wildlife Biology Program and Department of Ecosystem Conservation and Sciences, \\ University of Montana, Missoula, MT}\textrm{\Large{, }}
\Large{Jacob \textsc{Rash}$^{3}$} 
\let\thefootnote\relax\footnote{\baselineskip=10pt 3, 
Trout Unlimited, Arlington, VA}\textrm{\Large{, }}
\textrm{\Large{and }}
\Large{Mevin B. \textsc{Hooten}$^{1}$} 
\end{center}

\begin{abstract}
Phenomena such as epidemiological processes, hydrologic systems, social platforms, utility services, and supply chains can be represented as topological networks. A central question about these networks concerns connectivity and the permeability of edges. Dyadic regression and related approaches have been proposed to identify network features associated with pairwise node-level differences. In high-dimensional settings, it is important to control the number of spuriously selected features. However, controlling the false discovery rate for dyadic outcomes is challenging because dependence among dyads invalidates classic asymptotic procedures and complicates standard data splitting and knockoff approaches. We propose a novel knockoff variable selection procedure that simulates synthetic features directly on the topological network prior to constructing the augmented design matrix in dyadic space. Empirically, our method controls the false discovery rate for both node- and edge-level features. The Benjamini--Hochberg, Benjamini--Yekutieli, Storey Q-value, data-splitting, and standard knockoff procedures were all anticonservative. We applied our network knockoffs to assess the impassability of over $1000$ stream barriers in North Carolina for \textit{Salvelinus fontinalis}. Compared to data splitting and traditional knockoff approaches, our proposed approach selected a higher proportion of barriers previously assessed to impede fish movement.
\end{abstract}


\section{Introduction}

Network analyses have been increasingly adopted for the analysis of graph-structured data \citep{Goldenberg2010, Sengupta2025}. Various phenomena can be represented by a collection of nodes connected by a set of edges. A fundamental question concerns the existence of connections between nodes and various statistical procedures have been proposed to answer this. Inferring structure is especially challenging for high-dimensional networks, for which data have become more widely available \citep{Kolaczyk2014}. In this context, identifying all relevant connections is infeasible, and the burden shifts to limiting the number of features falsely concluded as important. 

A subset of false discovery methods specifically pertains to Gaussian graphical models \citep{Lauritzen1996}, which seek to determine the graph structure (i.e., the set of true edges) given the observed distribution of node-level features \citep{Liu2013, Fan2020, Li2021, Yu2021, Koka2024}. The central tenet of such models is that two nodes $i$ and $j$ are conditionally independent if there is no edge connecting them. Specifically, let $\bm{x}=(x_1,\dots,x_V)'$ be a vector of random variables observed on $V$ nodes. The true set of edges that connect the $V$ nodes are inferred using a combination of node-level regressions, $x_v|\bm{x}_{-v}$, paired with sparse optimization procedures specifically designed for graphical data \citep{Yuan2007, Friedman2008}. 

We considered a setting where the network is known and the goal is to identify associations between pairwise node-level differences and node- and edge-level features. In other words, we estimate the strength of connections among nodes given their graphical structure. For example, in a case study, we consider the impact of stream barriers on habitat connectivity for an aquatic species. The topology of a stream network is known but the degree to which barriers impact dependence among node-level responses is not. In a simulation study, we also considered the context in which edge conductance is a function of covariates measured at adjacent nodes, which has also been applied to model spatially structured ecological networks \citep{Hanks2013, Hanks2017, Peterson2019}.

We developed network knockoffs, synthetic negative controls with graphical construction. The proposed methodology is an adaptation of the Model-X knockoff framework \citep{Candes2018} tailored for networks where the typical Gaussian approximation of knockoff variables may not be appropriate. Our contribution relates to the construction of knockoff variables and an empirical investigation of their performance and was motivated by previous studies that demonstrated the benefits of using prior biological \citep{Sesia2019, Sesia2021} or topological \citep{Fan2020, Li2021} information to construct knockoff variables. 
While a variety of methods have been proposed for inferring conductivity in networks, we limited our analysis to dyadic regression \citep{Kenny2020} because of its simple interpretation, computational ease, and wide application. Dyadic regression has been used to infer the influence of landscape features on habitat connectivity based on observed genetic differences \citep{Wang2013, Schwob2024, Schwob2025}. \cite{Warren2023} estimated transmission rates of pathogens using dyadic regression. Dyadic regression can also approximate gravity models \citep{Gopinath2014} that are used for modeling international relations and trade \citep{Hoff2005, Graham2020}. 

Multiple studies have motivated dyadic regression as a computationally efficient alternative to traditional mechanistic approaches \citep{Graham2020, Schwob2024}. In the context of variable selection, a phenomenological approach is not necessarily a drawback provided that the simplified model can still identify associations generated from a more complex one \citep{Sesia2019, VanEe2025}. In a simulation study, we assess the power of dyadic regression approaches to detect relevant associations in datasets generated from models based on the Laplacian matrix associated with the underlying network. Furthermore, a central advantage of the proposed network knockoff methodology is that it is compatible with mechanistic modeling frameworks, compared to asymptotic \citep{Benjamini1995, Benjamini2001, Storey2002}, data splitting \citep{Dai2023}, Gaussian mirror \citep{Xing2023}, and traditional knockoff \citep{Barber2015} procedures that are challenging to generalize beyond sparse generalized linear models. 

In what follows, we establish notation conventions and statistical approaches for dyadic regression. We provide a brief overview of the traditional Model-X knockoff approach before introducing our network-based extension. Section \ref{simulation_section} is split into two parts that empirically assess selective inference for node- (\ref{simulation_nodes}) and edge-level (\ref{binary_edge_simulation}) features, respectively. In Section \ref{stream_barriers_section}, we apply our approach to assess the impassability of over $1000$ stream barriers for \textit{Salvelinus fontinalis}. Section \ref{discussion_section} concludes with a discussion of our findings.

\section{Methods}\label{methods_section}

\subsection{Preliminaries}

We use the notation $\mathcal{G}=\left(\mathcal{V}, \mathcal{E}\right)$ to define a graph with nodes $\mathcal{V}=\{1,\dots,V\}$ and edges $\mathcal{E}=\{1,\dots,E\}$. We assumed the graph is fully connected (i.e., each node $v$ has at least one edge connecting it to another node). Let $\bm{Y}=\left(\bm{y}_1,\dots,\bm{y}_V\right)$ be a $n\times V$ matrix of $n$ observations measured at each node. We used the notation $\bm{X}$ to define a matrix of $p$ features, which could pertain to the nodes or edges of the graph and lower case light, lower case bold, and upper case bold symbols to represent scalars, vectors, and matrices, respectively. We defined $\tilde{y}_{ij}\equiv f(\bm{y}_i,\bm{y}_j)$ as a shorthand notation for an arbitrary transformation of the node-level responses. For example, $f(\cdot)$ may be the Euclidean distance between observations. Similarly, we used $\tilde{\bm{x}}_{ij}\equiv f_{ij}(\bm{X})$ to define transformed covariates pertaining to the dyadic outcome between nodes $i$ and $j$. For example, $\tilde{\bm{x}}_{ij}=|\bm{x}_i-\bm{x}_j|$ and $\tilde{\bm{x}}_{ij}
= \min\limits_{\mathcal{P}_{ij}} \sum_{e \in \mathcal{P}_{ij}} \exp\{\bm{x}_e\}$, where $e=1,\dots,E$ indexes edges and the minimum is with respect to all paths $\mathcal{P}_{ij}$ that connect nodes $i$ and $j$. We refer to $\bm{x}$ as features and their transformation $f(\bm{x})=\tilde{\bm{x}}$ as covariates.

We considered dyadic regression models of the form 
\begin{align}\label{dyadic_full}
    [\tilde{y}_{ij}|\bm{\beta}, \theta_i, \theta_j, \sigma^2] &= \mathcal{N}(\tilde{\bm{x}}_{ij}\bm{\beta}+\theta_i+\theta_j, \sigma^2),
\end{align}
where $\bm{\theta}=(\theta_1,\dots,\theta_V)'$ are node-level random effects with zero mean and variance $\tau^2$ and we use the bracket notation to define probability distributions \citep{Gelfand1990}. There are $N=\binom{V}{2}$ dyadic outcomes.  The parameter of interest is $\bm{\beta}$ which regulates how node- and edge-level features influence network connectivity. 

\subsection{Model-X Knockoffs}

In high-dimensional settings (i.e., large $p$), it may be of interest to identify a subset of covariates that influence connectivity with high confidence. In these contexts, it is prudent to control the false discovery rate (FDR). We considered that only a subset of the covariates are truly associated with the response. Let $S\subseteq\{1,\dots,p\}$ denote the set of features that influence dyadic responses. The true set $S$ is unobserved and must be estimated. A covariate, $
\tilde{\bm{x}}_l=(\tilde{x}_{12,l},\tilde{x}_{13,l},\dots,\tilde{x}_{(V-1)V,l})'$, is said to be conditionally independent of the response if $\tilde{\bm{y}}\perp\tilde{\bm{x}}_l|\tilde{\bm{X}}_{-l}$. In other words, the covariate $l$ holds no additional predictive capability for dyadic responses assuming the inclusion of all other covariates. From equation (\ref{dyadic_full}), we note that the conditional independence of $\tilde{\bm{x}}_l$ is equivalent to $\beta_l=0$.  

Let $\hat{S}$ denote the subset of selected covariates. The FDR \citep{Benjamini1995} is defined as 
\begin{align}
    \text{FDR}=\mathbb{E}\left(\frac{\#\{l:l\in\hat{S} \text{ and } \beta_l=0\}}{\#\{l:l\in\hat{S}\}\bigvee1}\right).
\end{align}
The Model-X knockoff procedure controls FDR in finite samples by constructing a set of synthetic variables that satisfy two properties. Given a family of random variables $\bm{x}=(x_1,\dots,x_{p})'$, a Model-X knockoff, $\ddot{\bm{x}}=(\ddot{x}_1,\dots,\ddot{x}_{p})'$, satisfies 
\begin{enumerate}
    \item for any subset $S\subset\{1,\dots,p\}$, $\left(\bm{x},\ddot{\bm{x}}\right)_{\text{swap}(S)}\stackrel{d}{=}\left(\bm{x},\ddot{\bm{x}}\right)$,
    \item $\ddot{\bm{x}}\perp \!\!\! \perp\bm{y} | \bm{x}$,
\end{enumerate}
where $\stackrel{d}{=}$ denotes equality in distribution and $\left(\bm{x},\ddot{\bm{x}}\right)_{\text{swap}(S)}$ is obtained by swapping the variables $x_l$ and $\ddot{x}_l$ for all $l\in S$. Following \cite{Candes2018}, henceforth, we refer to criterion 1 and 2 as the exchangeability and nullity of knockoffs, respectively. 

The knockoff procedure can be applied to any variable selection method provided the original variables and knockoffs are treated symmetrically. The knockoff statistic is calculated from $w_l=g(x_l,\ddot{x}_l)$ where $g$ is any antisymmetric function, where large positive and negative values provide evidence for true and false discoveries, respectively. For example, $g$ could be the difference in magnitude of effects from a LASSO regression model \citep{Tibshirani1996} fit to the original and knockoff variable jointly (i.e., $g(x_l,\ddot{x}_l)=|\beta_l|-|\ddot{\beta}_l|$). The selection set $\hat{S}$ is obtained from choosing all variables with knockoff statistic exceeding $T$. \cite{Barber2015} showed the optimal selection criterion is given by 
\begin{align}\label{selection_threshold}
   T=\text{min}\left\{t\ge0:\frac{1+\#\{j:w_l\le -t\}}{\#\{j:w_l\ge t\}\vee1}\le q\right\},
\end{align}
where $q$ is the targeted false discovery proportion. 

The challenge with implementing Model-X knockoffs is that the distribution of features is generally unknown. \cite{Candes2018} suggested a second-order approximation of the features based on a Gaussian distribution, $\bm{x}\sim\mathcal{N}(\bm{\mu}, \bm{\Sigma})$. The joint distribution of original and knockoff features is specified as
\begin{align}\label{ModelXKnockoffs_gaussian}
    (\bm{x},\ddot{\bm{x}})\sim\mathcal{N}((\bm{\mu}',\bm{\mu}')', \bm{H})\text{, where } \bm{H}=\begin{pmatrix}
\bm{\Sigma} & \bm{\Sigma}-\text{diag}(\bm{s}) \\
\bm{\Sigma}-\text{diag}(\bm{s}) & \bm{\Sigma} 
\end{pmatrix},
\end{align}
and $\text{diag}(\bm{s})$ is any diagonal matrix selected in such a way that the joint covariance matrix is positive definite. 

A drawback of the Model-X knockoff procedure is that the selection procedure is based on one stochastic realization of the synthetic variables. Consequently, different instances of the knockoff procedure can result in discordant selection sets. To alleviate this issue, \cite{Ren2023} proposed aggregating results from $K$ replications of the knockoff procedure within a stability selection framework \citep{Meinshausen2010, Shah2013}. The derandomization procedure implies an upper bound for the expected number of false discoveries but does not control the FDR directly. 

\cite{Ren2024} adapted the derandomization procedure to preserve FDR control in finite samples using e-values \citep{Vovk2021}. Their first step involved applying the knockoff procedure to $K$ independently generated synthetic variables, with a target false discovery proportion of $\alpha_{\text{kn}}$. Using equation (\ref{selection_threshold}) to choose $T^{(k)}$, a selection set $\hat{S}_{\text{kn}}^{(k)}=\{l:w_l^{(k)}>T_l^{(k)}\}$ is chosen for $k=1,\dots,K$. For $l=1,\dots,p$, let 
\begin{align}
   e_l^{(k)}=p\cdot\frac{\mathcal{I}\{w_l^{(k)}\ge T^{(k)}\}}{1+\sum_{l=1}^{p}\mathcal{I}\{w_l^{(k)}\le -T^{(k)}\}},
\end{align}
where $\mathcal{I}\{A\}\in(0,1)$ is the indicator function for an event $A$. The collection of e-values is averaged as follows,
\begin{align}
   e_l^{\text{avg}}=\frac{1}{K}\sum_{k=1}^{K}e_l^{(k)}.
\end{align}
\cite{Ren2024} showed that applying the e-BH procedure \citep{Wang2022} to $\bm{e}^{\text{avg}}=(e_1^{\text{avg}},\dots,e_p^{\text{avg}})$ controls the false discovery rate at level $\alpha_{\text{ebh}}$. Note that $\alpha_{\text{kn}}$ may be different than $\alpha_{\text{ebh}}$ and only $\alpha_{\text{ebh}}$ controls the ultimate FDR of the derandomization procedure. The quantity $\alpha_{\text{kn}}$ can be viewed as a tuning parameter with $\alpha_{\text{kn}}<\alpha_{\text{ebh}}$ generally yielding greater power. 

\subsection{Network Knockoffs}

Note that the construction in equation (\ref{ModelXKnockoffs_gaussian}) describes the covariance between covariates but treats the entries within $\tilde{\bm{x}_l}$ independently. Because the observations of $\bm{x}_l$ for $l=1,\dots,p$ are distributed on a graph, the more relevant form of dependence may be among observations of the same feature as opposed to dependence across features. We construct knockoff variables that leverage prior information about the graph structure. While the exact model specification will be context dependent, we generally consider conditional autoregressive models and extensions of the form
\begin{align}\label{car_model}
    \bm{x}_{l}\sim\mathcal{N}\left(\bm{\mu}_{l}, \zeta^2_{l}\left(\text{diag}\{\bm{A}\bm{1}\}-\rho_{l}\bm{A}\right)^{-1}\right),
\end{align}
for $l=1,\dots,p$, where $\bm{A}$ is the node adjacency matrix associated with graph $\mathcal{G}$ and $\text{diag}\{\bm{A}\bm{1}\}$ denotes a diagonal matrix of its row sums. Motivations for modeling network data using discrete spatial models are discussed by \cite{Browning2017crime} and \cite{VerHoef2018}. Even if the $p$ features are independent, the dyadic covariates are likely to be correlated as a result of the shared graphical structure. Our approach involves first estimating the mean, variance, and range parameters from equation (\ref{car_model}), then simulating $p$ graphical knockoffs using 
\begin{align}\label{car_knockoff}
    \ddot{\bm{x}}_{l}\sim\mathcal{N}\left(\hat{\bm{\mu}}_{l}, \hat{\zeta}^2_{l}\left(\text{diag}\{\bm{A}\bm{1}\}-\hat{\rho}_{l}\bm{A}\right)^{-1}\right).
\end{align}
Because we have assumed independence, the simulated knockoffs satisfy exchangeability.  

\begin{figure}[tb]
\centering
\includegraphics[scale=0.7]{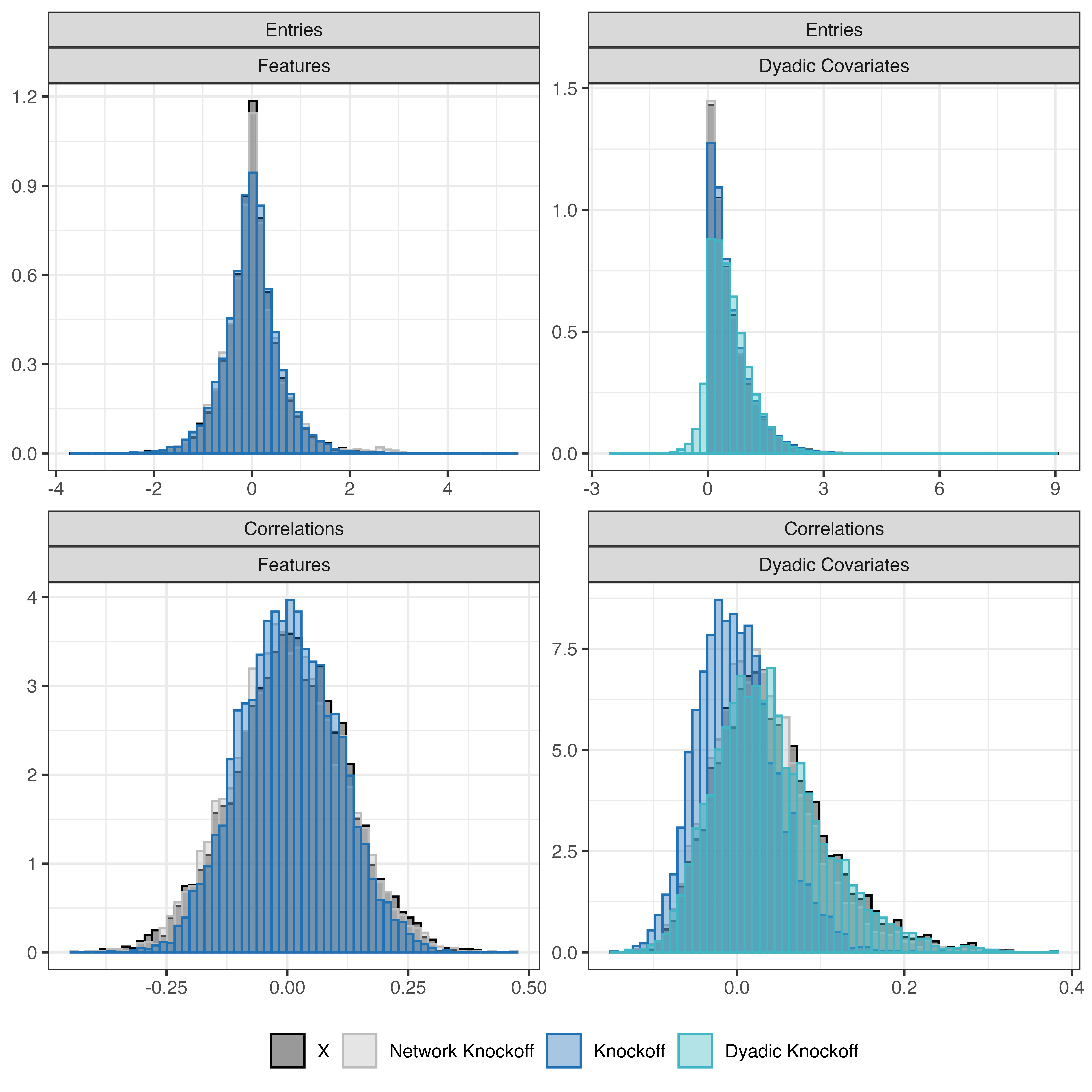} 
\caption{Histograms of entries and pairwise correlations for Gaussian graphical features, $\bm{X}$, Model-X knockoffs, and network knockoffs. Pairwise distances are calculated as the Euclidean distances between all nodes (i.e., $\tilde{x}_{ij}=(x_i-x_j)^2$). Knockoffs were generated by applying second-order knockoffs to the feature matrix $\bm{X}$. Dyadic Knockoffs were generated by applying second-order knockoffs to the transformed dyadic covariate matrix $\tilde{\bm{X}}$.}
\label{dyadic_histograms}
\end{figure}

Figure \ref{dyadic_histograms} shows the distribution of node-level features (left subplot) and resulting dyadic covariates (right subplot) based on pairwise Euclidean distances along with several knockoff variables. Second-order Gaussian knockoffs as implemented in the standard Model-X approach could be applied either to the node-level features $\bm{X}\in\mathbb{R}^{V\times p}$ or transformed dyadic covariate matrix $\tilde{\bm{X}}\in\mathbb{R}^{N\times p}$. In the presence of positive graphical dependence as implied by equation (\ref{car_model}), the transformed dyadic covariates will be positively correlated with small entries for proximate dyadic outcomes and large entries for distant dyadic outcomes. This correlation structure is not captured by the second-order node knockoffs (blue histogram, bottom-right subplot), which results in a distribution of correlations centered near zero. The second-order dyadic knockoffs (teal) provide the correct distribution of pairwise correlations but result in implausible negative dyadic distances (top-right subplot). The distributions associated with our network knockoffs (gray), which were generated from the estimated CAR model (\ref{car_knockoff}), are nearly identical to the distributions of $\bm{X}$ and $\tilde{\bm{X}}$. 

Like Model-X knockoffs, many standard FDR methods are ill-suited for dyadic regression because of the dependence of observations arising from repeated dyadic observations on the same node and graph structure. Nevertheless, a variety of classical FDR approaches have been successfully adapted for dependent data using mixed-modeling and residualization procedures. For example, in the context of genome-wide association studies, mixed-models are employed to account for familial relatedness and population structure \citep{Yu2006, Kang2010, Zhou2012, Sul2018}. A related set of approaches employ sequential procedures to first decorrelate the response and then perform variable selection \citep{Price2006, Price2010}, although these steps can be performed simultaneously using restricted regression \citep{VanEe2025}. Hence, while the dyadic framework violates the standard setup for most conditional and marginal testing approaches, these procedures may still control the FDR with the right pre-processing steps. We investigate this empirically in the following simulation study. 

\section{Simulation Study}\label{simulation_section}

We performed two simulation studies to empirically validate our approach and compare it with existing methods. We considered a variety of competitors for FDR control that can be classified into marginal and conditional testing approaches. The marginal testing approaches included Benjamini--Hochberg \citep{Benjamini1995}, Benjamini--Yekutieli \citep{Benjamini2001}, and Storey's Q-value procedures \citep{Storey2002}. Each of these procedures relies on p-values calculated from independent model fits or tests of association. The Benjamini--Hochberg (BH) procedure is the least conservative and may result in inflated false discovery rates for certain forms of dependence. Storey's Q-value (Qval) can accommodate weak dependence, and the Benjamini--Yekutieli (BY) is valid for any type of dependence but is also the most conservative. We qualify these general considerations noting that all three procedures have been shown to control FDR for certain covariance structures \citep{Stevens2017}. 

The conditional testing approaches included our proposed network knockoffs, second-order Model-X knockoffs \citep{Candes2018}, and data splitting \citep{Dai2023}. The second-order Model-X knockoffs were generated using \texttt{R} package \texttt{knockoff} \citep{knockoff2022} using the approximate semidefinite programming construction. We adopted the high-dimensional data splitting procedure and fit LASSO regression and ordinary least squares (OLS) to the equal halves of data. The vector of mirror statistics was calculated as $\bm{M}=\text{sign}(\hat{\bm{\beta}}_{\text{OLS}}\times\hat{\bm{\beta}}_{\text{LASSO}})(|\hat{\bm{\beta}}_{\text{OLS}}|+| \hat{\bm{\beta}}_{\text{LASSO}}|)$, which is the optimal functional form under certain conditions \citep{Ke2024}. We split datasets by dyadic observation, which violates the assumption of independence. We considered splitting the datasets by nodes, but this approach resulted in much lower power because of the large decrease in dyadic observations, ${\binom{V}{2}}=N>>2\times{\binom{V/2}{2}}$. For brevity, we do not report results. We also implemented the multiple data splitting procedure described by \cite{Dai2023} for $50$ randomly sampled partitions.  

All approaches were applied to residual dyadic observations $\tilde{r}_{ij}=\tilde{y}_{ij}-\hat{y}_{ij}$ taken from one of two models. The first was a simple linear regression model $\tilde{y}_{ij}\sim\mathcal{N}(\mu+\xi\log(d_{ij}), \varsigma^2)$ meant to account for the variation in $\tilde{y}_{ij}$ based on the structure of the graph and physical distance between nodes, $d_{ij}$. The second was a mixed-model $\tilde{y}_{ij}\sim\mathcal{N}(\mu+\xi\log(d_{ij})+\theta_i+\theta_j, \varsigma^2)$ that additionally corrected for the non-independence of dyadic observations because of shared nodes. The first and second model were fit using OLS and penalized maximum likelihood \citep{Nocedal1980} within the probabilistic programming language \texttt{Stan} \citep{Carpenter2017}. Although a subset of the competing approaches could estimate these effects jointly with the dyadic covariates, measuring performance against the same set of residuals enhanced comparability. We refer to the residuals resulting from fits to the linear model and mixed-model, respectively, as LM and MM.

\subsection{Gaussian Node Features}\label{simulation_nodes}

For the first study, we considered dyadic outcomes simulated directly from model  (\ref{dyadic_full}) with an additional linear predictor, $\xi\log(d_{ij})$, for physical distance. We assumed $\theta \sim \mathcal{N}(0, \tau^2)$. We generated small-world networks, which have been observed in a variety of fields \citep{Newman2001, Guimera2005, Bassett2017}, using the Watts--Strogatz model \citep{Watts1998} as implemented in the \texttt{R} software package \texttt{igraph} \citep{igraph2026}. A simple rejection sampler was used to ensure each simulated graph was fully connected. We assigned random weights to each edge using a standard log-normal distribution. The physical distances $d_{ij}$ were the sum of these weights based on the shortest path, $\mathcal{P}_{ij}$, connecting nodes $i$ and $j$.  

We generated $l=1,\dots,p$ features from a conditional autoregressive process,
\begin{align}
    \bm{x}_{l} \sim \mathcal{N}\left(\gamma_0+\gamma_1\bm{z}_l, \zeta^2\left(\text{diag}(\{\bm{A}\bm{1}\}-\rho\bm{A}\right)^{-1}\right),
\end{align}
where $\bm{z}_l$ was another node-level covariate. We modeled the dyadic outcomes as a function of the absolute difference of node-level covariates (i.e., $\tilde{x}_{ij,l}=|x_{i,l}-x_{j,l}|$), which has been used in several applications \citep{Hanks2013, Hanks2017, Schwob2024, Schwob2025}. The motivation behind this particular transformation is that processes may flow more easily between nodes with similar characteristics. For example, an invasive species may more easily travel along invasion corridors with habitats similar to the ones it has adapted to \citep{Gamba2025}. 

We applied a second-order Gaussian (i.e., matching mean and covariance) approximation to the dyadic design matrix $\tilde{\bm{X}}=\left(\tilde{\bm{x}}_{1},\dots,\tilde{\bm{x}}_{p}\right)$, where $\tilde{\bm{x}}_{l}$ denotes a column vector of all $N$ dyadic observations of transformed node-level covariate $l$. This approach does not account for the underlying graph structure, and the generated knockoffs may violate exchangeability. We created our network knockoffs by fitting $p$ independent spatial autoregressive models, and drawing one realization from the estimated multivariate distribution $\mathcal{N}\left(\hat{\gamma_0}+\hat{\gamma_1}\bm{z}_l, \hat{\zeta^2}\left(\text{diag}(\{\bm{A}\bm{1}\}-\hat{\rho}\bm{A}\right)^{-1}\right)$. Maximum likelihood estimates and predictions were obtained using \texttt{R} package \texttt{spmodel} \citep{Dumelle2023}. 

We simulated $100$ datasets from model (\ref{dyadic_full}) for all $27$ combinations of the following hyperparameters $p=50,100,200$, $\bm{\beta}(S)=0.25,0.5,1$, and $\tau^2=0,1,2$. This simulation design was selected to assess differences in power and FDR control based on variation in the number of features, signal strength, and network dependence, respectively. The other relevant hyperparameters were fixed $\sigma^2=1$, $\zeta^2=1$, $\gamma_0=0$, $\gamma_1=1$, and $V=30$. The correlation parameter $\rho$ was drawn from a $\text{Uniform}(0,1)$ distribution for each dataset. Note that $V=30$ ensured there were at least $200$ dyadic observations in each split for calculating $\hat{\bm{\beta}}_{\text{OLS}}$ $p=50,100,200$. The set of true features, $S$, was generated from a $p$-length vector of Bernoulli random variables with independent probability of success $0.3$. 

\begin{figure}[!htb]
\centering
\includegraphics[scale=0.56]{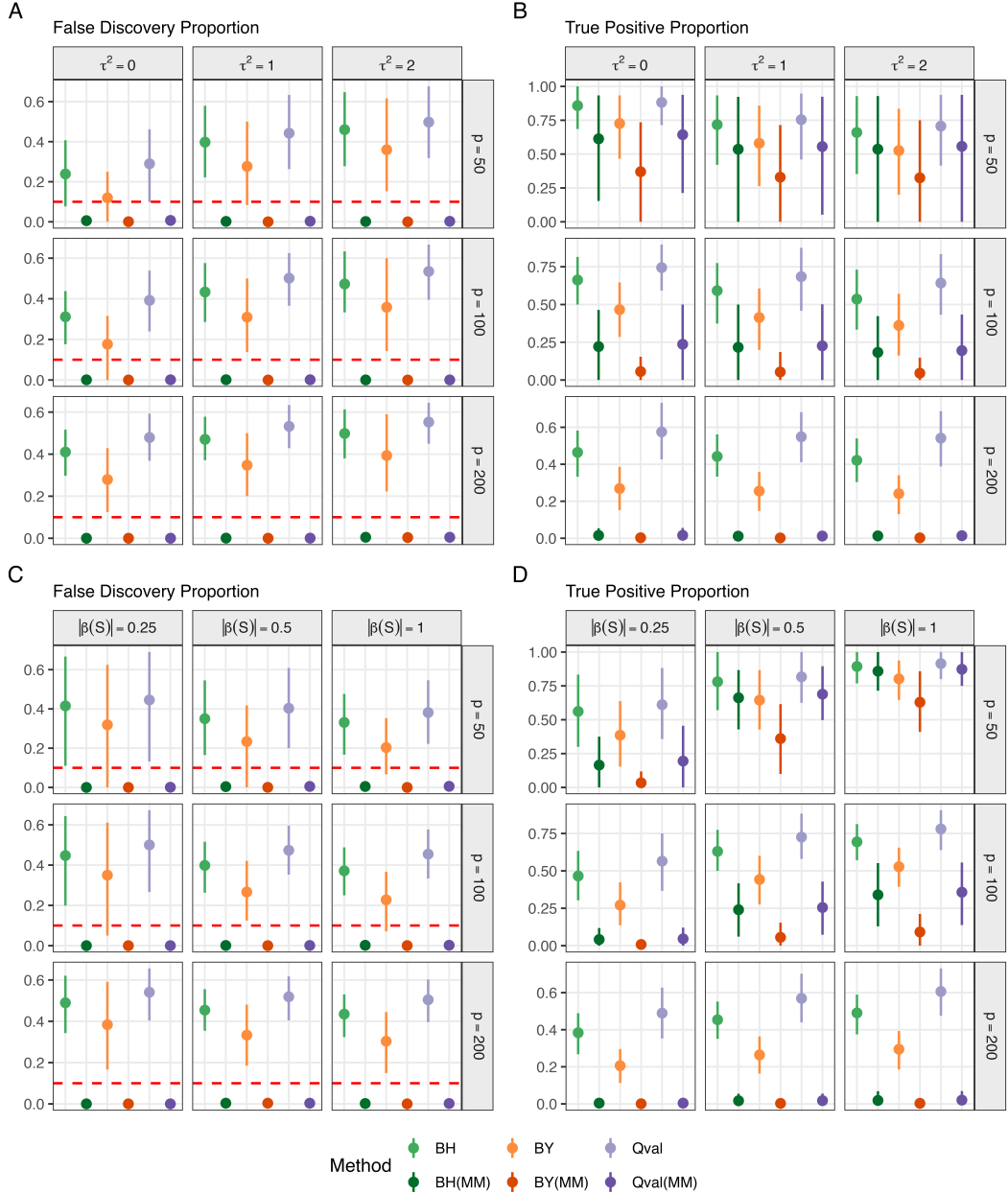} 
\caption{False discovery and true positive proportions for marginal testing approaches. Procedures shown include the Benjamini--Hochberg (BH), Benjamini--Yekutieli (BY), and Storey's Q-value (Qval) procedure. The targeted false discovery rate of $q=0.10$ is given by the dashed red line. The notation `(MM)' indicates procedures fit to the residuals of a mixed-model with node-level random effects. Subplots A and B are pooled across signals $|\beta(S)|=0.25, 0.5, 1$. Subplots C and D are pooled across dyadic dependence levels $\tau^2=0.25, 0.5, 1$. Line segments depict pooled mean, $10$th, and $90$th quantiles. The false discovery rate was less than $0.01$ for all marginal testing approaches applied to the mixed-model residuals in subplots A and C. The mean true positive proportions for BH(MM), BY(MM), and Qval(MM) were all less than $0.02$ for $p=200$ in subplots B and D.} 
\label{marg_dyadic}
\end{figure}

None of the marginal testing approaches controlled the FDR for the LM residuals (Figure \ref{marg_dyadic}, subplots A and C). The false discovery rate was lowest for the BY procedure but still elevated, especially for datasets simulated with more dyadic dependence (i.e., larger values of $\tau^2$, subplot A). In contrast, all approaches were conservative for the MM residuals and resulted in very few false discoveries. The BY(MM) procedure had the least power (Figure \ref{marg_dyadic}, subplots B and D). The Qval(MM) and BH(MM) procedures resulted in nearly identical power which decreased as the number of features increased. 

\begin{figure}[!htb]
\centering
\includegraphics[scale=0.56]{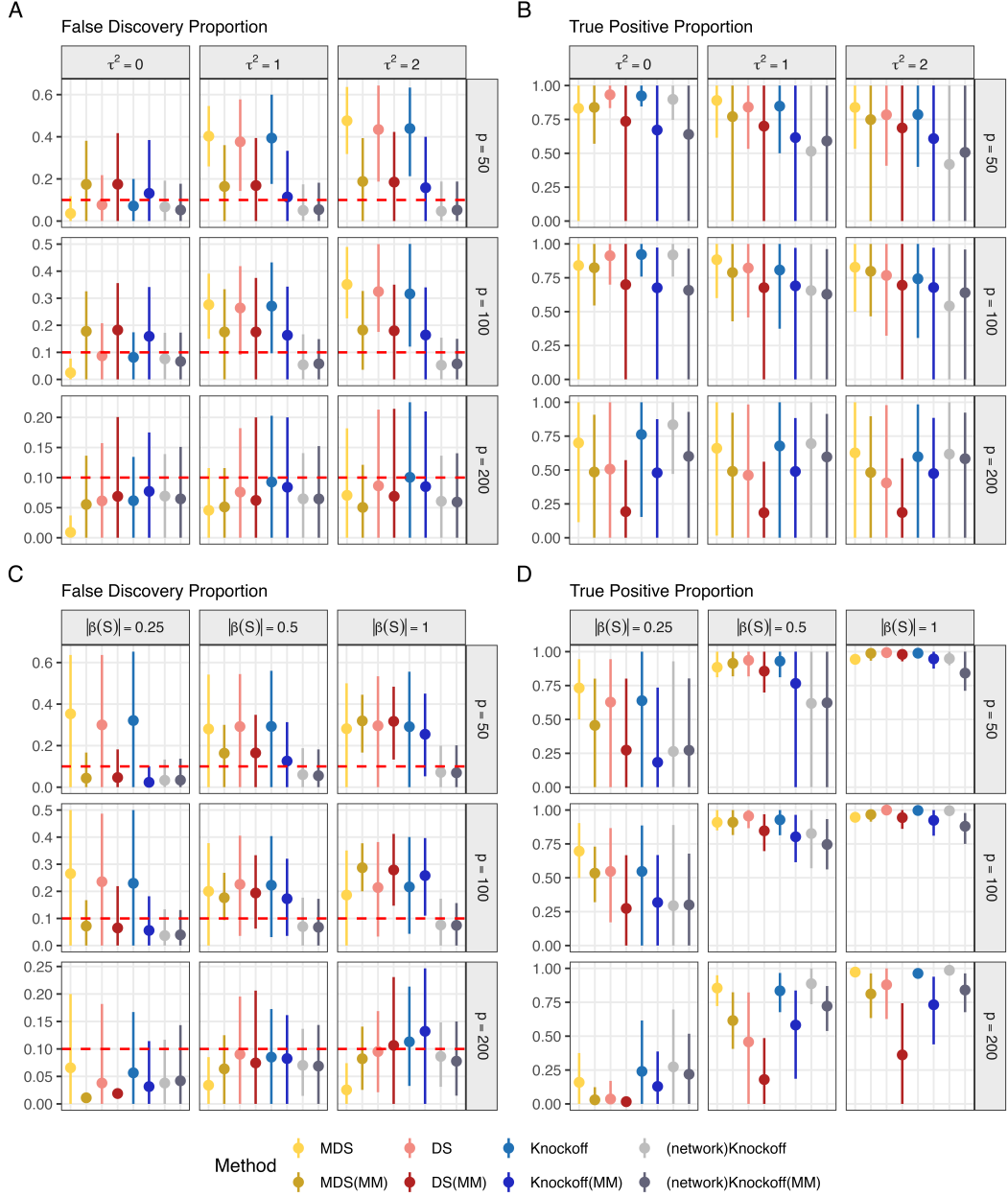} 
\caption{False discovery and true positive proportions for conditional testing approaches. Procedures shown include our proposed network knockoffs, second-order Model-X knockoffs, data splitting, and multiple data splitting. The targeted false discovery rate of $q=0.10$ is given by the dashed red line. The notation `(MM)' indicates procedures fit to the residuals of a mixed-model with node-level random effects. Subplots A and B are pooled across signals $|\beta(S)|=0.25, 0.5, 1$. Subplots C and D are pooled across dyadic dependence levels $\tau^2=0.25, 0.5, 1$. Line segments depict pooled mean, $10$th, and $90$th quantiles.} 
\label{cond_dyadic}
\end{figure}

Only our network knockoffs controlled FDR across the range of simulated datasets considered (Figure \ref{cond_dyadic}, subplots A and C). The data splitting approaches resulted in elevated FDRs for datasets simulated with more dyadic dependence, greater signals, and fewer features (i.e., the four panels in the top right of subplots A and C). Accounting for dyadic dependence using the mixed model lowered the FDR, but the mean FDR still exceeded the targeted $0.10$ level for the data splitting and traditional knockoff approaches. For the simulation settings in which all methods controlled the FDR, $\tau^2=0$ (leftmost column of subplot A) and $p=200$ (bottom row of subplot A), the network knockoffs also had low variation in false discovery proportion, comparable to multiple data splitting. For this same set of panels, power was generally highest and comparable for multiple data spitting, second-order knockoffs, and our network knockoffs (bottom row of subplot B). Applying the procedures to the mixed-model residuals generally yielded lower power (subplots B and D). Multiple data splitting applied to the mixed-model residuals resulted in fewer false discoveries but greater power than our network knockoffs for the smallest signal strength $|\bm{\beta}(S)|=0.25$ (leftmost panels of subplots C and D). Comparing to Figure \ref{marg_dyadic}, the conditional testing approaches achieved similar power to marginal testing for $p=50$ but outperformed them for $p=100$ and $p=200$.

\subsection{Binary Edge Features}\label{binary_edge_simulation}

In the second simulation study, we considered features measured on the edges of the graph within a mechanistic model for network connectivity. Graphs were generated using a preferential attachment algorithm \citep{Price1965, Barabasi1999}, which ensured $\mathcal{G}$ was fully connected and that there was only one path, $\mathcal{P}_{ij}$, connecting nodes $i$ and $j$. This formulation matches the structure of stream networks, an application we describe in the following section. A subset of graphs were also taken from physical stream networks observed in the eastern United States. We simulated a single binary edge feature $\bm{x}=\left(x_1,\dots,x_E\right)'$, where $x_e\in\{0,1\}$ indicates the presence of a barrier on edge $e\in\{1,\dots,E\}$. We use the terminology barrier because we simulated edges with $x_e=1$ to have less conductance than those with $x_e=0$. 

We simulated $n$ node-level observations from the model 
\begin{align}\label{laplacian_model}
    y_{\iota,v} \sim\mathcal{N}(\bm{0},(\bm{Q}\bm{Q}')^{-}+\sigma^2\bm{I}_V),
\end{align}
for $\iota=1,\dots,n$ and $v=1,\dots,V$, where $\bm{I}_V$ denotes the $V\times V$ identity matrix and $(\bm{Q}\bm{Q}')^{-}$ is a generalized inverse of $\bm{Q}\bm{Q}'$. The matrix $\bm{Q}$ is the Laplacian matrix with entries 
\begin{align}\label{laplacian}
    \bm{Q} = \begin{pmatrix}
\sum_j \alpha_{1j} & -\alpha_{12} & \cdots & -\alpha_{1V} \\
-\alpha_{21} & \sum_j \alpha_{2j} & \cdots & -\alpha_{2V} \\
\vdots & \vdots & \ddots & \vdots \\
-\alpha_{V1} & -\alpha_{V2} & \cdots & \sum_j \alpha_{Vj}
\end{pmatrix},
\end{align}
where $\alpha_{ij}$ is the conductance between nodes $i$ and $j$. Conductance was modeled using the piecewise function 
\begin{align}\label{conductance}
    \alpha_{ij} = 
        \begin{cases}
        \exp\{\beta_0+\beta_e x_{e(ij)}+d_{ij}\xi\} & \text{if there is an edge, $e(ij)$, connecting nodes $i$ and $j$} \\
        0 & \text{otherwise}
        \end{cases},
\end{align}
where $d_{ij}$ is the distance between nodes $i$ and $j$ or, equivalently in this context, the length of edge $e$. The Laplacian model (\ref{laplacian_model}) can be motivated by the limiting distribution of a random walk \citep{Hanks2017} and has been used to model the impact of barriers in ecological networks \citep{Hanks2013, White2020, Stack2026}.

While we simulated data from the mechanistic model described above, each FDR procedure was still applied to the dyadic regression model (\ref{dyadic_full}) for computational and comparison purposes. Note that, unlike the competing procedures, our network knockoffs could be incorporated directly into the mechanistic model described by \cite{Hanks2017}, albeit with greater computational burden because of the additional knockoff effects. The dyadic outcomes were calculated from the Euclidean distance based on all $n$ independent realizations of the process modeled by the Laplacian. The covariates $\tilde{\bm{x}}_{ij}=(\tilde{x}_{1},\dots,\tilde{x}_p)'$ indicated which of the $p=\sum_{e=1}^{E}x_e$ barriers were crossed on the path connecting nodes $i$ and $j$. Hence, even though we only have one graphical feature, we have $p$ dyadic covariates. As in the previous simulation study, we performed the same preprocessing step to remove spatial and node-level variation in the dyadic outcomes. 

The binary edge features were simulated using a spatial generalized linear mixed model (SGLMM) with exponential covariance function 
\begin{align}\label{SGLMM}
    x_e &\sim \text{Bernoulli}(\pi_e) \\
    \bm{\pi} &\sim \mathcal{N}\left(\gamma_0+\gamma_1\bm{z}_l, \zeta^2\exp\{-\bm{D}/\phi\}\right),
\end{align}
where $\bm{D}$ was the physical distance matrix between all edges and $\phi$ was the spatial range parameter, which was drawn from a uniform random variable bounded between the minimum and maximum distances observed divided by $3$. We let $\gamma_1=1$ and fixed $\gamma_0=\text{logit(0.2)}$, such that on average $20\%$ of edges had a barrier. Hence, the number of barriers was variable across simulations, but $\mathbb{E}\left(\frac{p}{V}\right)=0.2$ was fixed. The elements of $S$ (i.e., which barriers actually influence connectivity) were simulated independently from a vector of Bernoulli distributions with probability $0.3$. To generate the network knockoffs, $\ddot{\bm{x}}$, we first obtained parameter estimates $\hat{\gamma}_0$, $\hat{\gamma}_1$, $\hat{\zeta^2}$, and $\hat{\phi}$ by fitting the binary SGLMM, equation (\ref{SGLMM}), to the observed feature $\bm{x}$ using \texttt{R} package \texttt{spmodel}. We then obtained a sample from the conditional distribution $[\ddot{\bm{x}}|\hat{\gamma}_0, \hat{\gamma}_1, \hat{\zeta^2}, \hat{\phi}, \sum_{e=1}^{E}\ddot{x}_e=p]$ using a rejection sampling approach. 

We simulated $50$ datasets from model (\ref{laplacian_model}) for all $54$ combinations of the following hyperparameters $V=200,300,400$, $|\bm{\beta}(S)|=2,3,4$, $\zeta^2=0.5,1,2$, and graphs simulated using preferential attachment versus observed stream networks. For the observed graphs, we randomly drew a hydrological network with at least $V$ nodes. This simulation design was selected to assess differences in power and FDR control based on variation in sample size, signal strength, and feature estimation in addition to graph composition and sparsity. Note that for larger values of $\zeta^2$, it is more challenging to accurately estimate the distribution of edge features, which could potentially degrade the performance of our network knockoffs. Sparsity arises in the real networks because only a subset of nodes are observed. The other relevant hyperparameters were fixed $\sigma^2=1.0$, $\tau^2=0$, and $n=30$. 

The traditional Model-X knockoff approach has several drawbacks in this setting. Applying a second-order approximation to the feature matrix $\bm{X}$ is infeasible because only one realization of the binary edge features is observed. Applying the approximation to the dyadic design matrix $\tilde{\bm{X}}$ produces continuous-valued knockoffs, which are implausible. Our network knockoffs avoid both issues by generating synthetic barriers that preserve the binary structure and positive dependence of $\tilde{\bm{X}}$, as paths connecting more distant nodes tend to cross many of the same barriers.

\begin{figure}[!htb]
\centering
\includegraphics[scale=0.56]{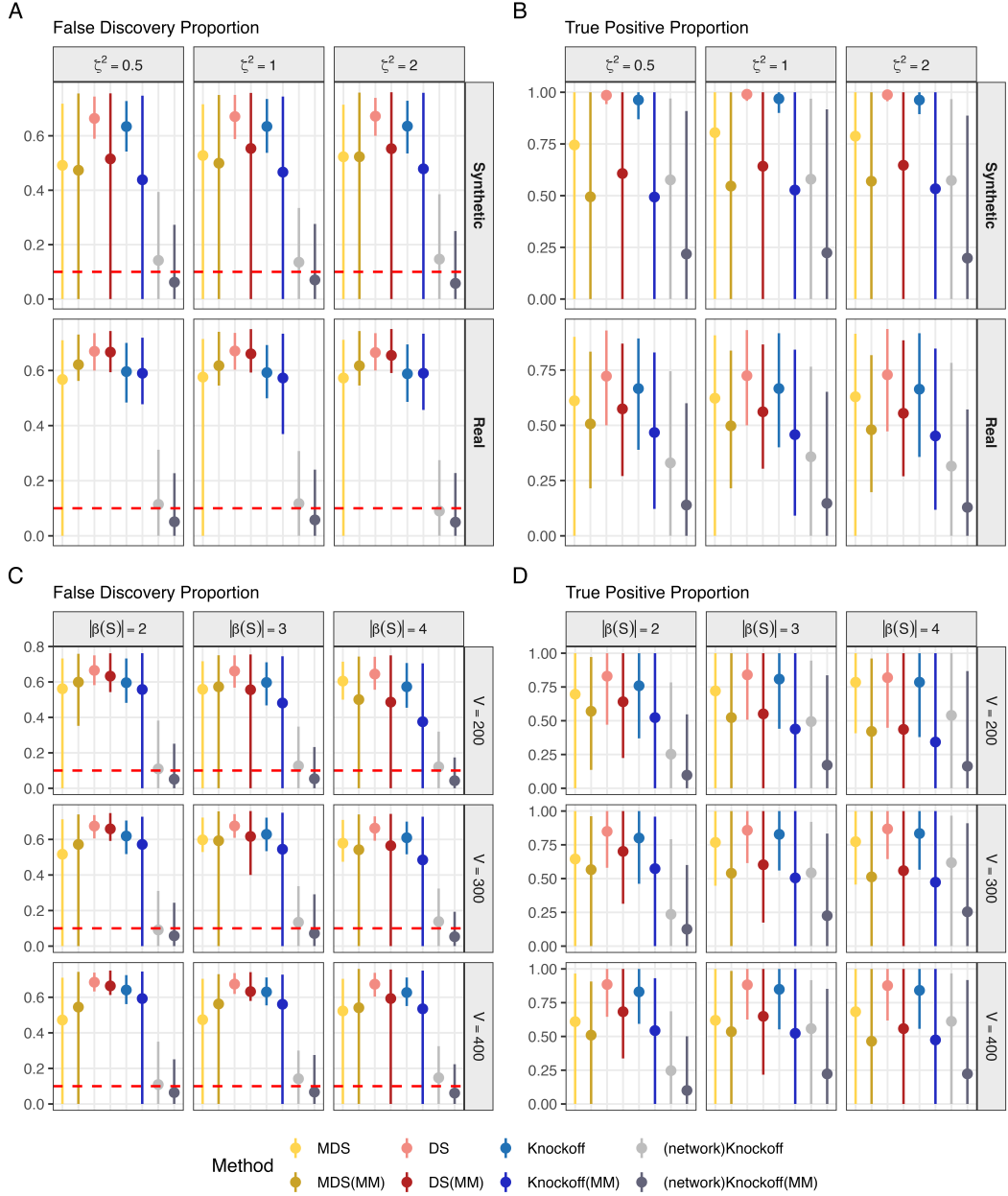} 
\caption{False discovery and true positive proportion for conditional testing approaches. Procedures shown include our proposed network knockoffs, second-order Model-X knockoffs, data splitting, and multiple data splitting. The targeted false discovery rate of $q=0.10$ is given by the dashed red line. The notation `(MM)' indicates procedures fit to the residuals of a mixed-model with node-level random effects. Subplots A and B are pooled across signals $|\beta(S)|=2, 3, 4$ and graph sizes $V=200,300,400$. Subplots C and D are pooled across graph type, preferential attachment versus observed stream networks and spatial variance of features $\zeta^2=0.5,1,2$. Line segments depict pooled mean, $10$th, and $90$th quantiles.} 
\label{cond_laplacian}
\end{figure}

All methods except our network knockoffs had severe FDR inflation (Figure \ref{cond_laplacian}, subplots A and C). The mean FDR for the data splitting, multiple data splitting, and second-order knockoff procedures, whether fit to the linear or mixed-model residuals, all exceeded $0.5$. The same was also true for all marginal methods, and for brevity, we do not report their results. The false discovery rate of our network knockoffs applied to the linear model residuals, was slightly elevated, $\overline{\text{FDR}}=0.12$ across all $2\text{,}700$ simulations. The empirical mean FDR met the targeted nominal $0.1$ level for $|\bm{\beta}(S)|=2$ (subplot C, leftmost column) but was elevated for the larger signals. The network knockoffs applied to the mixed-model residuals resulted in a false discovery rate of $0.06$ and were less powerful (subplots B and D). 

Figure \ref{cond_laplacian}, subplots A and B provide the performance of each conditional method split by graph types (synthetically generated via preferential attachment versus observed stream networks) and spatial variance of edge features $\zeta^2$. Another desirable property of methods which control FDR is low variance. Occasionally, our network knockoffs resulted in a false discovery proportion that exceeded $0.5$. Nonetheless, the $90$th quantile was less than $0.4$ for real stream networks (subplot A, top-row) and less than $0.3$ for networks simulated via preferential attachment (subplot A, bottom-row). Applying our network knockoffs to the mixed-model residuals decreased both the mean and variance of the false discovery proportion (subplot A) and lowered power (subplot B). 

Power was lower for real stream networks than for synthetic graphs (Figure \ref{cond_laplacian}, subplot B). Further investigation revealed this was not a function of graph composition but rather graph sparsity. Mean power was $0.43$ for $50$--$100\%$ of nodes observed, $0.14$ for $33$--$50\%$ observed, and $0.05$ for less than $33\%$ observed. As before, the network knockoff procedure applied to the mixed-model residuals was more conservative. Increasing the spatial variance, $\zeta^2$, did not appear to affect the false discovery or true positive proportions.  

\section{Effect of Stream Barriers on Habitat Connectivity for \textit{Salvelinus fontinalis}}\label{stream_barriers_section}

An important question in riverscape conservation relates to the impact of stream barriers on habitat connectivity \citep{ErHos2012, Branco2014, ErHos2019}. Larger patches bolster population size, genetic diversity, and dispersal and thereby facilitate conservation \citep{Hodgson2009, Liczner2024}. Consequently, various efforts have sought to improve patch sizes through the removal of stream barriers \citep{Mckay2017}. The drawbacks of incorrectly identifying (i.e., a false discovery) a habitat barrier are primarily related to a misallocation of resources to removal but could also extend to additional economic consequences related to water storage, irrigation, recreation, and flood prevention \citep{Mckay2017}. Variable selection and FDR control are essential for prioritization and limiting the removal or alteration (e.g., installing fish ladders or passages) of barriers that do not impede habitat connectivity \citep{Katopodis2005, Kemp2010, Noonan2012}.

Several methods have been proposed for assessing the influence of stream barriers on habitat connectivity. A collection of statistical approaches use genetic data to infer connectivity based on the pairwise fixation index distances, $F_{\text{ST}}$, within random walk models for spatially structured ecological networks \citep{Hanks2013, Hanks2017, Peterson2019}. Analogous to the dyadic approaches described, multiple regression can be applied directly to the genetic distance matrices to estimate associations with barrier data \citep{Balkenhol2009}. Using a discriminant analysis of principal components \citep{Jombart2010}, the effect of barriers on connectivity has also been inferred from the spatial arrangement of population clusters estimated with genetic software such as \texttt{STRUCTURE} \citep{Pritchard2000}. Barrier impassability can be assessed based on physical factors (e.g., height, pool depth, water flow) and morphometric characteristics of the species \citep{Gibson2005, Meixler2009, Meixler2021}.

We assessed the impassability of $1590$ barriers in the eastern United States for a native salmonid, \textit{Salvelinus fontinalis} (brook trout). Brook trout are economically valuable and an indicator species for high-quality coldwater streams \citep{Kanno2010}. While populations can persist in stream segments of less than one kilometer \citep{Letcher2007, Kanno2011a}, large habitat patches can buffer brook trout from the negative impacts of rising temperatures and competition from nonnative salmonid species by providing access to cold water refugia \citep{Petty2012}. The effect of habitat barriers on brook trout connectivity has previously been assessed \citep{White2020, Stack2026} but these analyses modeled barriers jointly, as opposed to an independent assessment of barrier impassability. To our knowledge, we present the first statistical method for assessing barrier impassability with exact FDR control. 

The Southeast Aquatic Resources Partnership (SARP) has mapped thousands of barriers in the United States to assess impediments to habitat connectivity and facilitate conservation planning \citep{Martin2019, SARP2025}. Each barrier is marked as either a dam or road-related barrier (culvert). The impassability of each barrier is classified as ``Complete barrier,''  ``No barrier,'' ``Partial passability,'' and ``Unknown.'' While these classifications are not specific to brook trout, it is reasonable to assume that dams inhibit movement more than culverts and that the complete and no barriers classifications are accurate. It is also reasonable to assume that partially passable barriers could fragment habitat connectivity as young-of-year trout have more limited jumping ability and swimming strength. Note that the SARP dataset does not include all anthropogenic barriers or any natural barriers (e.g., waterfalls and riffles), which also decrease habitat connectivity \citep{Kelson2015}.  

\begin{figure}[!htb]
\centering
\includegraphics[scale=0.35]{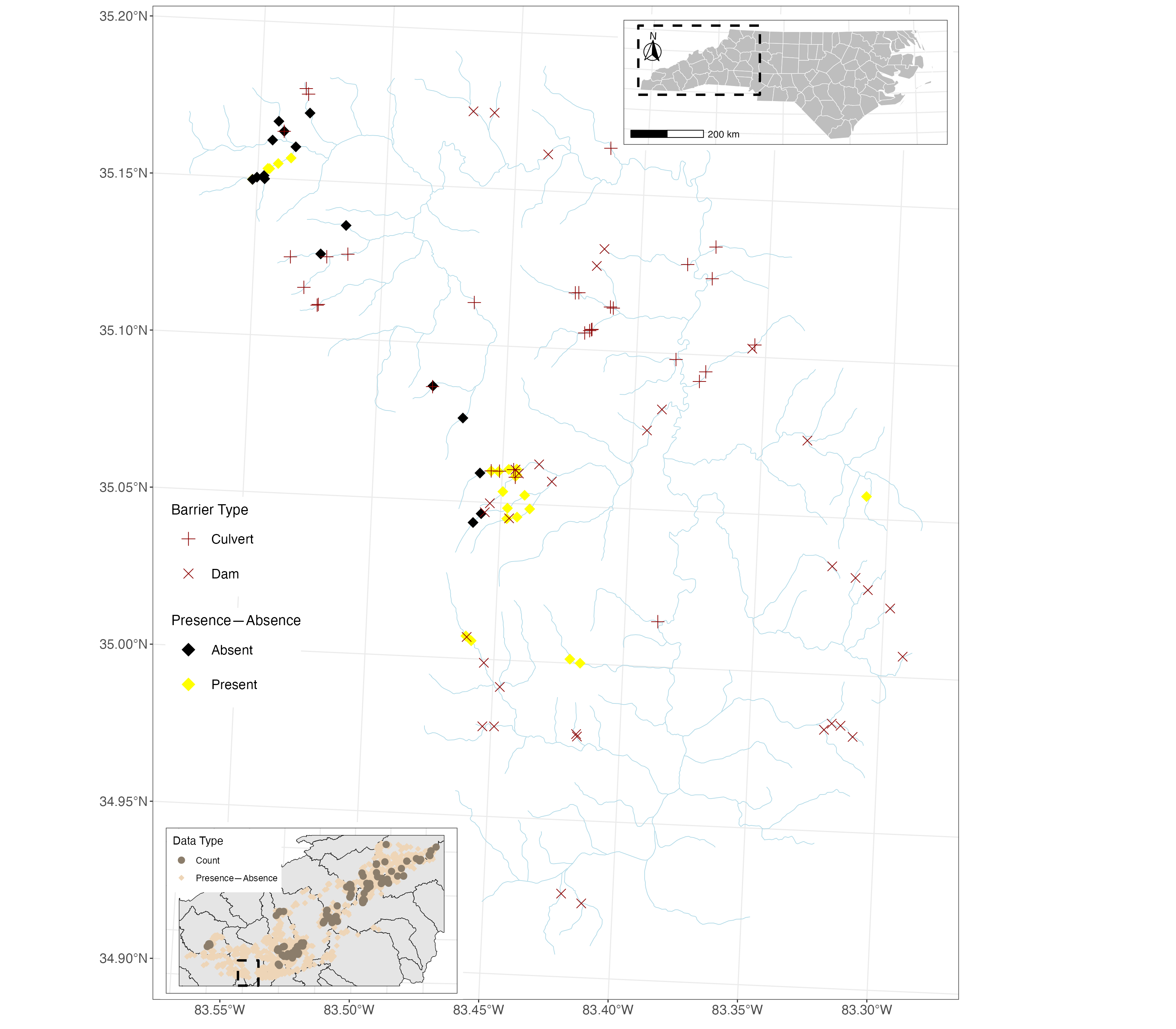} 
\caption{Map of data sources used for assessing impassability of stream barriers. Top-right inset shows the study area in relation to North Carolina. Bottom-left inset shows the locations of all $133$ electrofishing (count) and $1931$ presence--absence sites. Main plot is one of the 10 stream networks (blue lines) considered in our analysis.} 
\label{data_sources_north_carolina}
\end{figure}

Unlike the preceding approaches, which used genetic or morphometric data, we assessed habitat connectivity based on a latent spatial random effect inferred from an integrated species distribution model \citep{Pacifici2017, Schliep2018, Isaac2020, Simmonds2020, Watson2021} fit to electrofishing and presence--absence data within a data fusion approach \citep{Gelfand2025}. We provide a map of our spatial count, presence--absence, and barrier data in Figure \ref{data_sources_north_carolina}. We accounted for dependence in brook trout abundance across space and time by modeling the intensity surface as a log-linear combination of multiple effects 
\begin{align}\label{intensity_process}
\text{log}(\lambda_{i,t})=\log(a_{i,t})+\bm{x}'_{i,t}\bm{\beta}+\xi_{t}+\zeta_{w(\iota)}+\eta_{\iota}, 
\end{align}
where $a_{i,t}$ is the stream habitat area sampled at site $i$ in year $t$ in square meters (divided by $1000$ to improve numerical stability; \citeauthor{Lu2024a}, \citeyear{Lu2024a}, \citeyear{Lu2024b}), $\bm{x}_{i,t}$ is a vector of covariates (intercept included), $\xi_{t}$ is a temporal random effect, $\zeta_{w(i)}$ describes broad-scale spatial patterns related to the subbasin identity, $w$, of site $i$, and $\eta_{i}$ accounts for within-network spatial dependence based on hydrologic distance. The covariates included in $\bm{x}_{i,t}$ were informed by recent analyses \citep{Lu2024b, Valentine2024} and are described in \citep{VanEe2026}. 

For within stream network dependence, we used a tail-down model \citep{VerHoef2006, Peterson2010, VerHoef2010}, which has been suggested as a suitable spatial covariance function for species that can travel upstream. Specifically, we specified the following exponential covariance model 
\begin{align}\label{eta_equation}
    \bm{\eta}_{s} \sim \mathcal{N}\left(\bm{0}, {\sigma^2_{\eta}}_s\exp\left({-\bm{D}_s/\phi_{s}}\right)\right),
\end{align}
where $\bm{D}_{s}$ is the hydrologic distance matrix (i.e., the distance between all sites when movement is restricted to the stream network) and $\bm{\eta}_{s}$, a subvector of $\bm{\eta}$, is the spatial random effect for stream network $s=1,\dots,14$. From equation (\ref{intensity_process}), we can interpret the stream-level spatial effect $\eta_{i}$ as the residual density of fish after accounting for differences in survey area, habitat, watersheds, and temporal dynamics.

The tail-down model induces an assumption of isolation-by-distance, but using the posterior distribution of $\bm{\eta}$, we could infer more complex spatial dynamics. For example, holding all else equal, large differences in the spatial random effect between geographically proximate sites would suggest the barrier separating them negatively influences connectivity. Accordingly, we fit the following dyadic regression model to assess the impassability of each barrier 
\begin{align}
\tilde{y}_{ij}|\bm{\beta},\sigma^2\sim\mathcal{N}\left(\tilde{\bm{x}}'_{ij}\bm{\beta},\sigma^2\right),
\end{align}
where $\tilde{y}_{ij}=(\eta_i-\eta_j)^2$ and $\tilde{\bm{x}}_{ij}$ is a binary vector indicating which of the $p$ barriers were crossed on the path connecting sites $i$ and $j$. 

We implemented six different FDR control methods: data splitting, multiple data splitting, second-order knockoffs, derandomized second-order knockoffs \citep{Ren2023, Ren2024}, network knockoffs, and derandomized network knockoffs. We used $K=100$ replications for multiple data splitting and the derandomized knockoff approaches. For data splitting, we followed the same procedure described in Section \ref{simulation_section} and computed the mirror statistics from OLS and LASSO models fit to the two randomly split halves of dyadic outcomes. As in Section \ref{binary_edge_simulation}, we applied a second-order Gaussian approximation to the full dyadic design matrix, $\tilde{\bm{X}}$, and generated knockoffs using approximate semidefinite programming. Some barriers within the stream network were never crossed by the paths connecting the observed nodes and $\tilde{\bm{X}}$ was not full-rank (i.e., columns corresponding to the uncrossed barriers have all $0$ entries). We applied the second-order approximation to the subset of barriers that were crossed, leaving the columns corresponding to uncrossed barriers unchanged in the generated dyadic knockoff matrix. 

We fit a binary SGLMM with logistic link function to all $14$ stream networks to estimate the probability each stream segment had a barrier. The model included fixed effects for Strahler stream order \citep{Strahler1957}, mean stream gradient, the natural logarithm of the stream's catchment (drainage) area, mean annual flow, and the fraction of the upstream catchment area that is covered by lakes, ponds, or other still bodies of water. All five covariates were scaled to have zero mean and unit variance. We used an exponential covariance function based on the Euclidean distance between barriers because their distribution is primarily influenced by topography and anthropogenic factors that cluster geographically. We calculated the Brier skill score as $BSS=1-\frac{BS}{BS_{null}}$, where $BS_{null}=\sum_{e=1}^{E}(x_e-\bar{x})^2/E$ for each network. The mean skill score was $0.25$ and varied from $0.11$--$0.41$ indicating moderate improvements in predictive performance. The mean Brier score \citep{Brier1950} was $0.08$ and varied from $0.04$--$0.12$. We generated the network knockoffs using the same rejection sampler described in Section \ref{binary_edge_simulation}. 

Ten of the $14$ stream networks fit with the integrated species distribution model had at least one stream barrier assessed by SARP. We applied each of the FDR methods to the networks independently. One could consider pooling knockoff and mirror statistics, but this would control FDR globally, not per stream network. From a conservation perspective, one watershed or stream network will often be selected for barrier removal projects based on habitat suitability, population size, and genetic diversity. Hence, barrier removal prioritization and FDR control is needed within, not across, networks.   

\begin{figure}[!htb]
\centering
\includegraphics[scale=0.46]{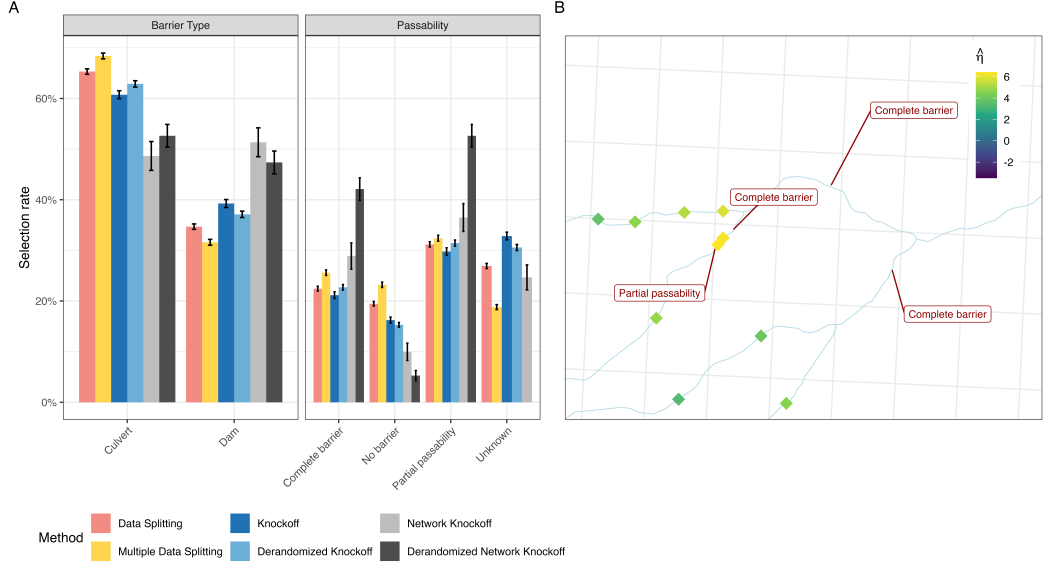} 
\caption{Subplot A: Selection rates of false discovery control methods. For data splitting, knockoff, and network knockoff, proportions are calculated from the pooled selections of all $K=100$ replications. The $95\%$ error bars were calculated from the asymptotic Wald interval of sample proportions. The targeted false discovery rate was $0.1$ for data splitting, knockoff, and network knockoff but $0.29$ for multiple data splitting, derandomized knockoff, and derandomized network knockoff. The Southeast Aquatic Resource Partnership defines two barrier types: dams and assessed road-related barriers (culverts). The impassability of each barrier is classified as either complete barrier, no barrier, partial passability, or unknown. \\ Subplot B: Close up of four selected barriers on stream network plotted in Figure \ref{data_sources_north_carolina}. Filled diamonds indicate posterior mean of estimated spatial random effect (equation \ref{eta_equation}) from integrated species distribution model for $10$ presence--absence sites.} 
\label{barrier_selection_proportions}
\end{figure}

We targeted a nominal false discovery rate of $0.1$. At this level, no barriers were selected by the derandomized knockoff approaches. We increased the nominal rate to $0.29$ to obtain a non-empty selection set, $\hat{S}$. The selection rates of each method are reported in subplot A of Figure \ref{barrier_selection_proportions}. Among the methods that depend on a single replication (i.e., data splitting, knockoff, and network knockoff), our network approach selected a higher proportion of dam related barriers, which are more likely to impede fish movement than culverts. Our network approach also had greater selection rates for complete and partially passable barriers and lower rates for non-barriers than data splitting and the second-order knockoff approach. Derandomization further improved the performance of our network knockoffs resulting in a higher selection rate of complete barriers while decreasing selection of non-barriers. Derandomization degraded performance of the second-order knockoffs. 

Averaged over all $K=100$ replications, our network approach selected $\approx12$ barriers across all stream networks versus $\approx149$ and $\approx303$ barriers selected per replication with second-order knockoffs and data splitting, respectively. Similarly, $228$ and $254$ barriers were selected using the derandomized second-order knockoff filter and multiple data splitting, respectively, versus $19$ using our derandomized network approach. These differences corroborate the anticonservative tendencies of the second-order knockoff and data splitting approaches reported in Section \ref{binary_edge_simulation}. 

We plot four barriers selected using our network knockoffs and their derandomized extension in subplot B of Figure \ref{barrier_selection_proportions}. The four barriers were selected in $35\%$ of replications of our network knockoffs which was the highest selection frequency among all barriers. The two barriers on the right-hand side of subplot B appear to separate moderate values of the spatial random effect in the bottom stream segments from relatively higher values in upper segments. These barriers also separate these sites from the larger stream network shown in Figure \ref{data_sources_north_carolina}. The complete and partial barriers on the left-hand side isolate two sites with very high estimated values of the spatial random effect from five nearby sites with relatively lower estimates.

\section{Discussion}\label{discussion_section}

Through a series of empirical investigations and a case study, we have shown that classical and state-of-the-art selective inference approaches fail to control the false discovery rate in network analyses. The central issue with multiple testing and data splitting procedures is that dyadic observations are not independent, and network structure is challenging to account for using typically applied mixed effect models. Network analyses also present several challenges to the implementation of standard Model-X knockoff approaches. Dyadic transformations of graphical features result in a complex distribution of possibly discrete covariates that is difficult to capture using second-order Gaussian approximations. 

We developed a knockoff implementation that leverages prior information about the network to create more meaningful synthetic negative controls. Our network knockoffs were the only method shown to control FDR in all scenarios and generally resulted in the highest power in contexts where multiple methods controlled the FDR. Another advantage of our network knockoffs is that they generally lend themselves to a physical interpretation. For example, in our analysis of stream barriers, Section \ref{stream_barriers_section}, our network approach generated pseudo-barriers for matched comparison. The pseudo-barriers could be used to control FDR in mechanistic modeling approaches that build the covariance among nodes from the Laplacian matrix \citep{Hanks2013, Hanks2017}. In this context, our network knockoffs can also generate new unidentifiable dyadic covariates in the form of pseudo-barriers that are never crossed by the collection of paths connecting the observed nodes. Such pseudo-barriers cannot be captured by standard Model-X knockoff approaches. 

One practical challenge associated with the analysis of stream barriers relates to the precision of geographic information systems. The coordinates of brook trout survey sites, location of stream barriers, and stream network layers were obtained from the North Carolina Wildlife Resources Commission, SARP, and National Hydrology Dataset Plus. Building a spatial stream network from these datasets involved snapping the survey sites and stream barrier locations to the stream edges, and this step can lead to topological errors in our analyses of stream networks. For example, some sites may have been located upstream of barriers that they are actually downstream of. Hence, the barriers selected from our analysis should be interpreted with caution. Future work could incorporate uncertainty in network topology directly into the modeling framework, which would further improve the reliability of barrier impassability assessments. Regardless, our analysis validates the usefulness of the network knockoff approach given the relatively higher selection proportion for complete and partially passable barriers.  
 
While we primarily motivated our approach in an ecological context, network knockoffs have applications in many other fields. In each case, the ability to limit false discoveries while maintaining power is essential for responsible inference on complex networked systems. For example, in epidemiology our approach could be used to identify which individual behaviors influence epidemic thresholds \citep{Leitch2019}. Likewise, binary network knockoffs could be used to detect broken pipes or electricity leakage in utility systems \citep{Watts1998}. Network knockoffs could also be applied to determine the factors which influence the flow of processes in social networks \citep{Price1965, Newman2001, Hoff2002, Browning2017, Larsen2024}.

\noindent \textbf{Acknowledgments}: This project was funded by a Multistate Conservation Grant F25AP00123 from the U.S. Fish and Wildlife Service and jointly administered with the Association of Fish and Wildlife Agencies.

\baselineskip=10pt
\bibliographystyle{apalike}
\bibliography{bibliography.bib}

\end{document}